\documentstyle[12pt,moriond,epsfig]{article}
\begin{document}
\heading{STELLAR DISTRIBUTIONS AND NIR COLOURS OF NORMAL GALAXIES}

\makeatletter
\renewcommand{\@makefnmark}{\mbox{\ }}
\makeatother

\footnote{Invited Review to apper in {\it
``Extragalactic Astronomy in the Infrared''}, eds.\ G.\ A.\ Mamon,
Trinh Xu\^an Thu\^an, \& J.\ Tr\^an Thanh V\^an, Editions
Fronti\`eres, Gif-sur-Yvette}

\author{R.F. Peletier $^{1}$, R. de Grijs $^{1}$} {$^{1}$ Kapteyn 
Astronomical Institute, Groningen, The Netherlands} 

\begin{moriondabstract}
We discuss some results of a morphological study of edge-on galaxies, based
on optical and especially near-infrared surface photometry. We find that the
vertical surface brightness distributions of galaxies are fitted very well
by  exponential profiles, 
much better than by isothermal distributions. We find that in general 
the vertical scale height increases when going outward. 
This increase is strong for early-type spiral galaxies and very small for late
types. We argue that it can be due to the presence of thick discs with scale 
lengths larger than the galaxy's main disc. 
Finally we discuss the colour-magnitude relation in $I-K$ for spiral galaxies.
We find that it is a tight relation, for which the scatter is similar to the 
observational uncertainties, with a steeper slope than for elliptical galaxies.
\end{moriondabstract}

\section{Introduction}

To learn more about the processes responsible for the formation and evolution of
galaxies one can either observe galaxies at high redshift to directly measure 
the evolution, or study nearby galaxies in more detail, to look for remnants
of the formation process. In this paper we do the latter, and discuss 
some aspects of the morphology and stellar populations of spiral galaxies.
This subject is not new. About 20 years ago the morphology of spiral
galaxies was studied extensively by van der Kruit \& Searle
\cite{vanderKruit81a,vanderKruit81b,vanderKruit82a,vanderKruit82b}.
At that time knowledge about stellar populations had come mainly from 
optical and near-infrared colour studies (e.g. \cite{Visvanathan77,Aaronson78}
and others). In the meantime, however, much better 
instruments and detectors have become available, requiring these problems
to be attacked again. Here (Section 2) we will discuss a morphological study 
in $B$, $I$ 
and $K$ of a large sample of edge-on spirals. The infrared array observations
allow us to look through the dust, so that the vertical profiles near the 
symmetry plane of the galaxy can be studied much better than before. Away from 
the plane our deep observations allow us to address the question whether 
thick discs are common in spiral galaxies.

Although optical and near-infrared colours of galaxies 
have been studied frequently,
their interpretation has been difficult as a result of extinction by dust. 
Only now infrared array observations have made it possible to obtain 
high-resolution optical and infrared maps, allowing people to choose areas
that are not affected by dust, and in this way really study the stellar
populations. In Section 3 we  will discuss one of these dustfree colour-
magnitude (CM) diagrams and investigate its implications. 

\section{Morphology of discs and bulges}

\subsection{Vertical profiles near the plane}

Van der Kruit \& Searle
\cite{vanderKruit81a,vanderKruit81b,vanderKruit82a,vanderKruit82b} 
found that vertical profiles 
of spiral galaxies resemble self-gravitating isothermal sheets with the
following light distribution:

\begin{equation}
K(z) = K_0 \hbox{ sech}^2 (z/z_0)
\end{equation}

where K(z) is the observed vertical distribution, and z$_0$ the vertical
scale parameter. 
At large $z$ this distribution goes asymptotically to an exponential 
distribution. Their result was based on photographic observations of
edge-on galaxies, which were severely affected by extinction. Later, more
observations, especially in the near-infrared, became available (e.g.
\cite{Wainscoat89,Aoki91}), showing that
for the nearby galaxies IC~2531 and NGC~891 much better fits were obtained with 
purely exponential functions. Up to now this problem has not been studied for
a large sample of edge-on galaxies. The problem with the exponential
distribution is that it not understood physically, and that it implies 
very low velocity dispersions in the plane of the galaxy \cite{vanderKruit88}.
Seeing both the exponential and the isothermal distributions as extremes,
he proposed to use a family of density laws:

\begin{equation}
\label{family.eq}
K(z) = 2^{-2/n} K_0 \hbox{ sech}^{2/n} (nz/2z_0) , \qquad (n > 0)
\end{equation}

For $n$=1 we have the sech$^2$ law (isothermal sheet), 
while for $n$=$\infty$
we have the exponential distribution. In Fig.~1a a family of these denisty laws
is shown.

\begin{figure}
\mbox{\epsfysize=9truecm \epsfbox{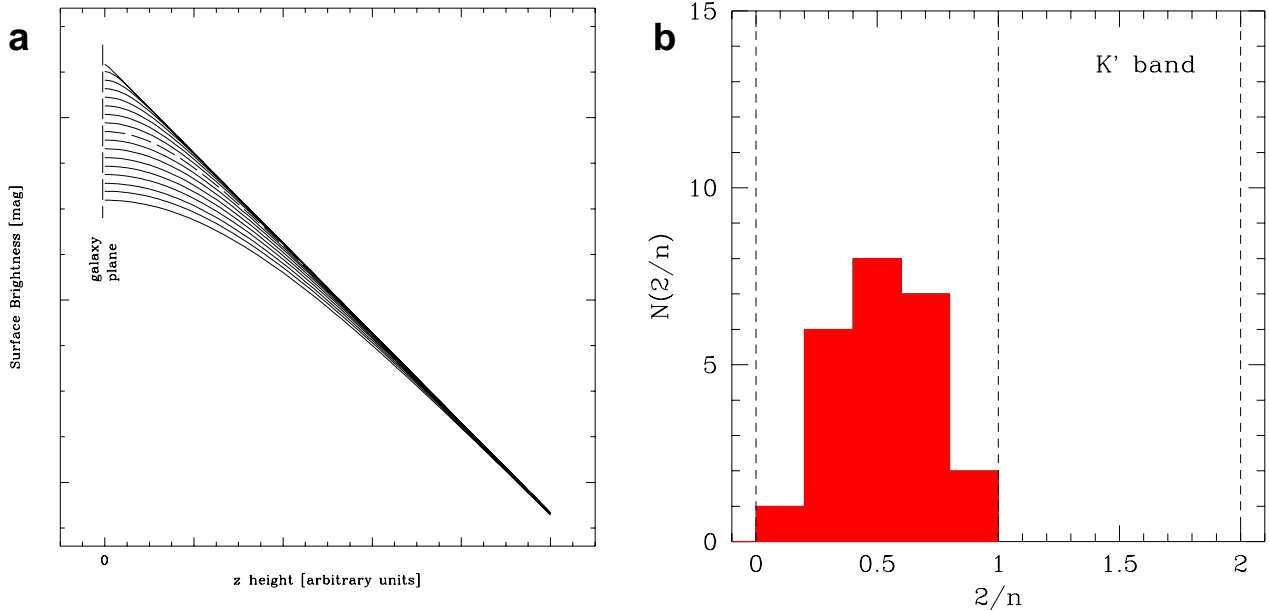}}
\caption{a: The family of density laws, with the isothermal (2/$n$=1) and 
the exponential (2/$n$=0) distributions as two extremes. The stepsize in
$n$ is 0.125; for clarity the sech($z$) model (2/$n$=1) is shown with dashed
lines. b: Histogram of average values of (2/$n$) obtained from the
extinction-corrected $K'$-band observations.}
\end{figure}

To determine for which $n$ the best fits of eqn. (\ref{family.eq}) could be made 
to real spiral galaxies, we have observed a sample of 24 galaxies in $K'$
\cite{Wainscoat92} (calibrated to $K$), 
$I$ and $B$. It is a subset of a complete diameter-limited sample of edge-on
galaxies with inclinations larger than 87$^{\rm o}$, comprising galaxy types
from S0 to Sd. Details about the observations are given in  \cite{deGrijs97b}. 
The observations have been taken at
ESO, La Silla, from 1993 to 1996. The infrared observations were taken with
IRAC2b on the 2.2m, while the optical observations came from the 1.5m Danish
and the 0.9m Dutch telescope. The observations were taken under photometric
conditions, and calibrated using standard stars, giving an accuracy of 
$\approx$0.08 mag in $K'$ and 0.04 mag in $I$ and $B$. The seeing ranged between
1'' and 1.5''. In Fig.~2 we plot some vertical profiles through the center
in $K'$ and $I$. 
\begin{figure}
\mbox{\epsfysize=12truecm \epsfbox{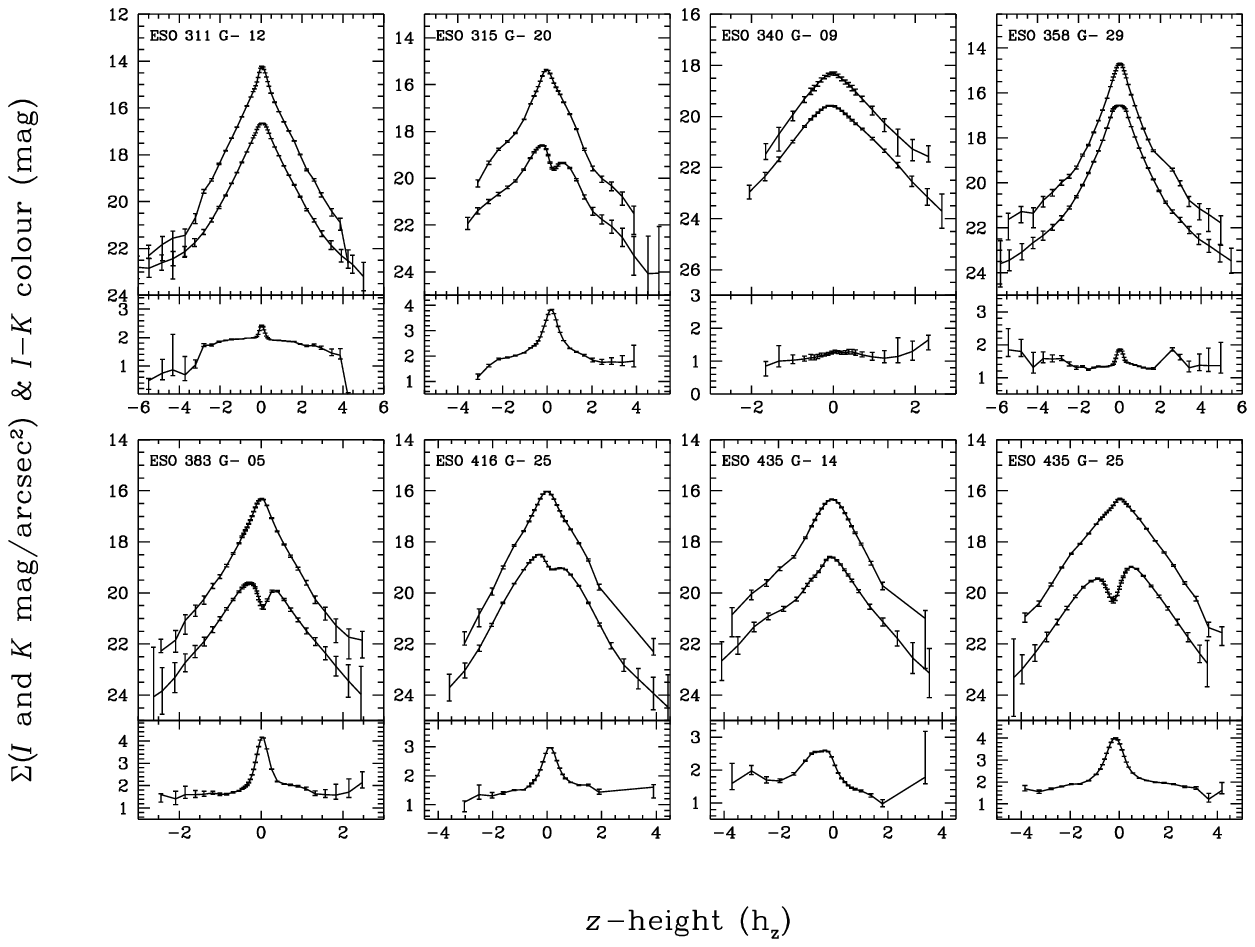}}
\caption{Central $I$ and $K'$ band vertical profiles of 8 galaxies, and
their corresponding $I-K$ colours.}
\end{figure}
It is clear that the profiles in the $I$-band near the 
symmetry plane are still significantly affected by extinction, while the
symmetry in $K'$ shows that extinction here barely affects the profiles.
In the bottom panels the $I-K'$ profiles give a good idea of the amount of 
extinction near the mid-plane, assuming that $I-K'$ is constant across the 
profile. This assumption is justified to a large extend 
by the fact that in the area where
$I-K$ is a smooth function of position the change in colour is small, and 
in the central regions, where a lot of dust extinction is observed, the 
changes are large and the profiles often asymmetric. 
Moreover, stellar population models
of old stellar populations predict very small changes in $I-K'$ (see e.g.
\cite{Vazdekis96}). For each profile we determined a constant,
dust-free colour, used it to determine E$_{I-K}$ at $z$=0, and corrected our $K'$
using the Galactic extinction law \cite{Rieke85} for residual
extinction. After this we fitted eqn. (\ref{family.eq}) to all the corrected $K'$
profiles, at various
radial distances in the galaxy. In general the fits were acceptable, 
and in Fig.~1b we give the histogram of average $n$-values per galaxy. The
radial distributions of $n$ are given in Fig.~3. For signal-to-noise reasons
we have binned together the data for galaxies of similar morphological types.
\begin{figure}
\mbox{\epsfysize=9truecm \epsfbox{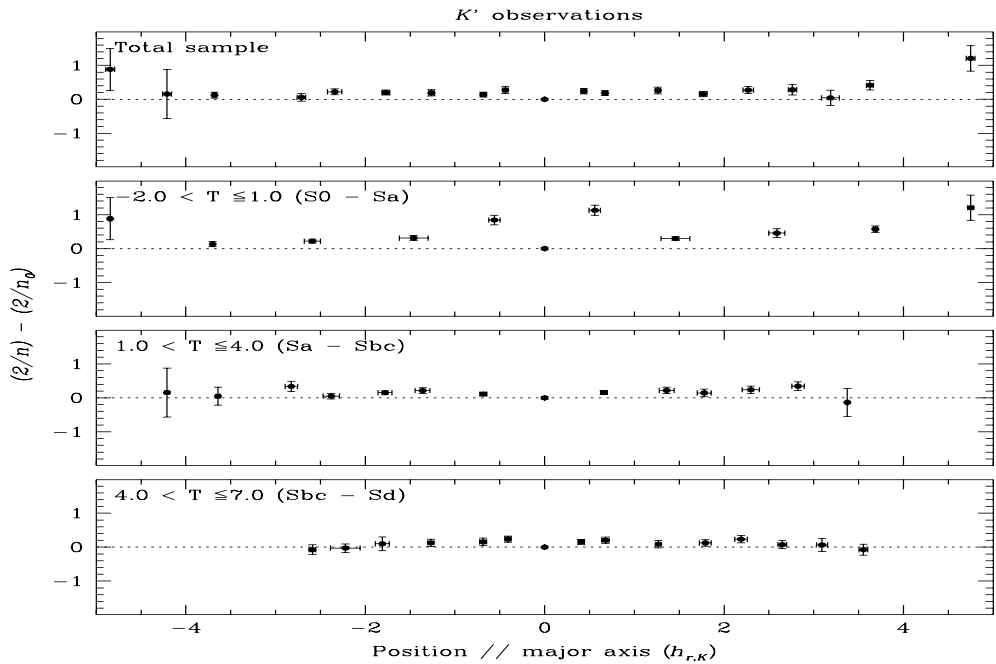}}
\caption{Averaged distributions of the sharpness parameter as a function
of position along the major axis, both for the total sample and for specific 
Hubble type bins.}
\end{figure}

We see that the observed vertical profiles of all spiral galaxies have shapes that 
lie between the exponential and sech($z$) distributions. Also, no variations as
a function of radius along the major axis are seen. 
We now consider what this implies for the three-dimensional, deprojected profiles.
If galaxies have an orientation of exactly 90 degrees, with everywhere the 
same vertical profile shape, projection along the line of sight leaves the 
profile shape invariant. However, at positions only a few degrees away from
edge-on, the profiles flatten near $z$=0, severely affecting the determination
of $n$. Simulations (described in \cite{deGrijs97b}) show that at 
$i$=87$^{\rm o}$ an exponential profile will look like a profile with 
(2/$n$) = 0.7 in projection, also somewhat dependent on position in the galaxy.
Because of the uncertainties in the determination of the inclinations, we can
only say that the average correction that we have to apply to (2/$n$) lies
between 0.3 and 0.5. Since we find (Fig.~1b) that $n$ = 0.54 $\pm$ 0.2
projected onto the plane of the sky, the deprojected profiles are probably 
very close to exponential ($n \approx$ 0.0-0.2).

This result agrees very well with observations of our own galaxy. From 
near-infrared maps of the sky Kent {\it et al.} \cite{Kent91} showed that the best
fits to our Galaxy could be obtained using an exponential distribution with
radially increasing scale length. Similarly Kuijken \& Gilmore \cite{Kuijken89} 
found
that the density profile of stars in the solar neighbourhood 
is also very close to exponential.
Currently there are no simple models leading to exponential vertical profiles.
Simulations of disc heating by spiral arms by Jenkins \& Binney \cite{Jenkins90} 
produce a remarkably isothermal distribution of the stars in the solar
neighbourhood. One way to make these distributions more peaked is to add
new bursts of star formation. New simulations will have to answer the question
of how much recent star formation for all spiral galaxies is implied by these
observations.

\subsection{Vertical profiles at large $z$-distances}

\begin{figure}
\mbox{\epsfysize=9truecm \epsfbox{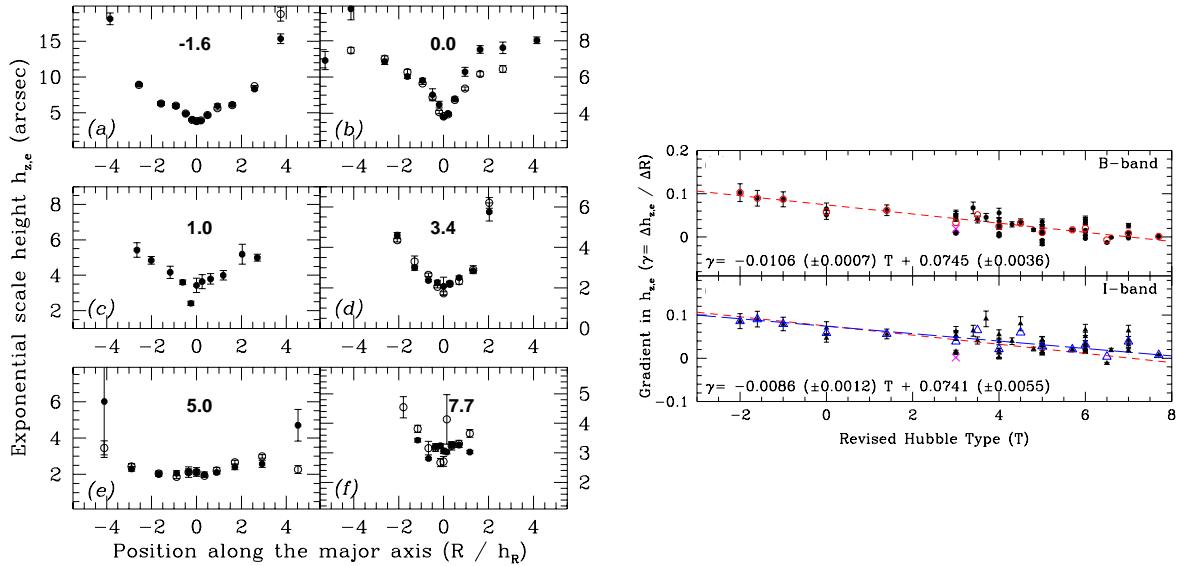}}
\caption{a: Examples of $I$-band scale height as a function of galactocentric
distance for 6 galaxies of our sample. Morphological types are shown.
Open and closed symbols represent data taken on both sides of the galaxy plane.
b: Disc scale height gradients as a function of Hubble type. The closed
symbols represent the data; the open symbols are type-averaged data points.
The result that we obtained using van der Kruit \& Searle \cite{vanderKruit81b}'s 
data for NGC 891 is indicated with crosses ($\times$). For comparison, 
the best fit obtained for the $B$-band data is also shown in the $I$-band
figure.} 
\end{figure}

In this subsection we analyze the outer parts of the vertical profiles, now
using the $I$-band observations, since the $K'$ band observations do not go
deep enough due to the high infrared sky background. Studies of edge-on
galaxies have generally shown that away from the central plane the light 
distribution can be fitted well by an exponential distribution, for which the 
radial variation of the scale height is typically within 3\%  of the mean
\cite{vanderKruit81a,vanderKruit81b,vanderKruit82a,vanderKruit82b,Shaw90,
Barnaby92}. The fact that the scale height is constant is 
hard to understand, since current theory predicts that the vertical
profile shape is determined by heating by molecular clouds, of which
the distribution changes considerably as a function of radius \cite{Jenkins92}. 
Other
mechanisms will have to be invoked (see \cite{deGrijs97a}).

For the sample described above we have analysed their vertical exponential
scale heights,
and their radial dependency. These results are shown for 6 galaxies of different
types in Fig.~4a. It shows that for every galaxy the scale length increases,
and that the increase is strongest for the earliest-type galaxies. In the area
of the bulge the scale heights are smallest. One might think that this effects
is caused by the fact that the bulge contaminates the fit to the disc. This is
incorrect - for example: Barnaby \& Thronson \cite{Barnaby92} find for NGC~5907 that the scale
height is lower between --100'' and +100'' than in the rest of the galaxy. 
They say that this due to their bulge contamination. However, as can be seen 
from their Fig.~4, between $\pm$50'' and $\pm$100'' the bulge contribution
is negligible. We find that in none of the objects bulge contamination is 
responsible for the increase of scale height with radius.

As a following step we have fitted the radial change in scale height 
between 2 and 4 $I$-band scale lengths. We find (Fig.~4b) that in all our
galaxies scale heights increase radially or remain constant. For morphological
types larger than 2 the change is very small, but for earlier types the
gradients are considerable.

What can we conclude? It has been known for a long time time that S0 galaxies
show the presence of a thick disc component, much stronger than the thick disc
in our Galaxy (\cite{Burstein79,Tsikoudi79}). They also see that this thick 
disc starts dominating the light in the outer parts, indicating that its 
scale {\it length} is larger than the scale length of the main disc. If one fits 
only one exponential to the vertical distribution, one will see an increase
of the scale height with radius. For the later types van der Kruit \& Searle 
\cite{vanderKruit81a} note
that the late type edge-on galaxies NGC~4244 and NGC~5907 have larger
scale heights in their outermost profiles. It seems that spiral galaxies in
general show the presence of thick discs with scale lengths larger than 
those of the main disc, and that the relative importance of the thick disc
goes down strongly towards later types. The origin of the thick discs
is not very clear. It seems that the presence of the thick disc is determined
completely by the morphological type, or maybe the gas fraction of the galaxy. 
The process that makes discs of S0 galaxies in general thicker than discs
of later type galaxies, e.g. interactions, would then also be responsible 
for the formation of the thick disc. At the surface brightness that we
are observing these galaxies are still very symmetric (see Fig.~4a), so 
we are not looking at the presence of warps here.

\subsection{Galactic bulges}

We end this section on the morphology of spiral galaxies by summarising 
some recent results for galactic bulges. For years the r$^{1/4}$ profile
\cite{deVaucouleurs48} has been extremely popular in modeling surface 
brightness of bulges (e.g. \cite{Kent86}), although often significant deviations
were observed, and no conclusive paper had appeared showing that this law
was fitting well in general for bulges. Presumably, people liked to use it,
to emphasise the evolutionary link with elliptical galaxies, which are
in general well fitted by it (e.g. \cite{Burkert93}). Andredakis \& Sanders
\cite{Andredakis94} showed that the photometric data of
Kent \cite{Kent86} could be fitted with exponential bulges 
just as well as with r$^{1/4}$ bulge profiles, with
much less internal scatter in the fitted parameters. Using higher quality
data, Andredakis {\it et al.} \cite{Andredakis95} subsequently showed that 
bulges of {\it early-type} spirals could be fitted in general well by the r$^{1/4}$
law, while {\it late-type} bulges looked much more exponential. Parametrising the
surface brightness law by:

\begin{equation}
K(r) = K_0 ~ \exp(-r/r_0)^{1/n}
\end{equation}  

they found that the fitted $n$ correlated very well with morphological type,
or with bulge-to-disc ratio. A likely cause of the relation is that the
interaction of bulge and disc had altered the bulge profile in the outer parts.
Recently however, the results from HST have again complicated this picture.
Apart from an exponential or r$^{1/4}$ bulge Phillips {\it et al.}
\cite{Phillips96} found
that many bulges of morphological type earlier than Sc also had an unresolved
stellar nucleus with very high surface brightness. More analysis 
is needed for us to understand these cusps and the interaction between
the cusps and the rest of the galaxy. 

\section{The colour-luminosity relation in spiral galaxies}

To understand the formation and evolution of galaxies it is very important
to study their scaling relations, i.e. relations between their fundamental
quantities like mass, diameter, surface brightness etc. For early-type 
galaxies there is a tight relation between colour and luminosity: 
Bower {\it et al.} \cite{Bower92a,Bower92b} determined that for the 
colour-magnitude 
relations in $U-V$ and in $V-K$ the scatter in both the Virgo and the 
Coma cluster was comparable to the observational scatter, and that the 
CM relations had the same slope. Early-type galaxies generally have very little
star formation and extinction by dust, so their colour directly reflects
their stellar populations. Bower {\it et al.} interpret the colour-magnitude 
relation as a relation between mass and metallicity, and in this interpretation
all early-type galaxies in Coma and Virgo must have formed at approximately
the same time. Ellis {\it et al.} \cite{Ellis96} recently also found tight CM
relations in 3 distant clusters, showing that the early-type galaxies in those 
clusters also must have formed at the same time, or are very old.

It is hard to derive similar constraints for spiral galaxies, mainly because of 
the fact that recent star formation and extinction are responsible for 
considerable scatter. It would however be very interesting to know more
about CM relations for spirals, especially to study their star formation
histories. When studying an infrared CM diagram one could 
study the older stellar populations in spirals, and find out whether 
they were formed at the same time as the early-type galaxies.
The optical-infrared CM relation was studied in detail by Tully {\it
et al.} \cite{Tully82}, who claimed that spiral galaxies had a different CM
relation than lenticulars, while both relations had very little scatter.
They claimed that the difference could be explained by the presence of
more star formation in late type spirals, affecting especially the blue
light, and hence the $B-H$ colour. Their result, except for the fact that
it was very hard to interpret, since they used the hybrid $B_T$ - $H_{-0.5}$
colour, was not confirmed by aperture photometry of Mobasher {\it et al.}
\cite{Mobasher86}, who found that early and late type galaxies had more or less
the same CM-relation in $B-K$ and $J-K$, although the scatter was much smaller
in $J-K$, but still considerable. Almost certainly Tully {\it et al.}'s 
data have not been corrected enough for extinction, since some of his 
S0 galaxies were redder than the reddest ellipticals. This is in disagreement
with Balcells \& Peletier \cite{Balcells94}, who showed that dustfree colours of bulges
of S0's are always bluer or have the same colour as ellipticals of the same
luminosity. These authors also don't reproduce the gap in colour that Tully 
{\it et al.} find between S0's and later type spirals.

Nowadays, using infrared arrays, it is possible to derive CM-relations 
with a much smaller scatter, by measuring for each galaxy the colour in 
areas where extinction is thought not to be important. We will
discuss here the $I-K$ vs. M$_K$ relation for a sample of spiral galaxies
consisting of two subsamples: 1) a sample of early-type spiral galaxies
with inclination $>$ 50$^{\rm o}$, where the bulge is unobscured by the 
disc (from \cite{Peletier97}). Of this sample we only use the 
galaxies of type Sa and earlier, to avoid dust-affected colours.  For this 
sample we took the bulge colour at 5'' from the center on the minor axis. 2) a 
sample of edge-on ($i$ $\ge$ 87$^{\rm o}$) galaxies, from de Grijs {\it et al.}
\cite{deGrijs97b}). Colour maps and profiles show that the colour in the vertically outer
parts is constant, and that the colour profiles are featureless and
symmetric compared to the other side of the galaxy. For this reason we assumed
that
all the extinction is concentrated  near the plane of the galaxy, and 
that the outer regions are dustfree. The colours were determined on both
side of the plane on the minor axis in the area where the colour is constant
(appr. 1-3 scale heights).

\begin{figure}
\mbox{\epsfysize=8.5truecm \epsfbox{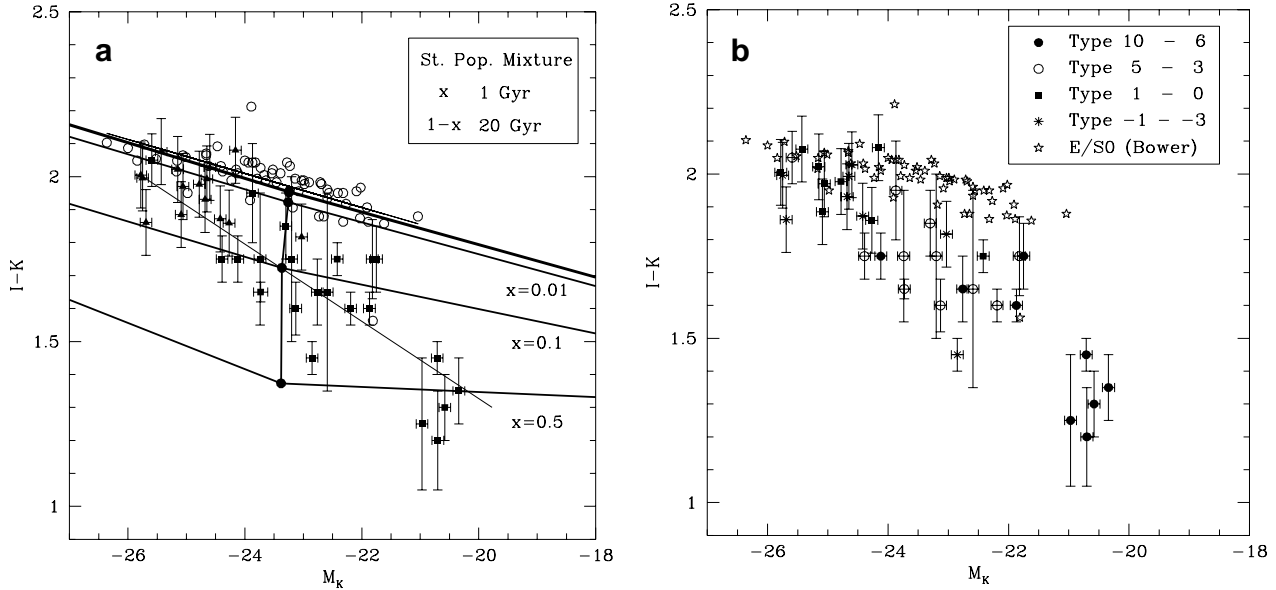}}
\caption{{\bf a:} The $I-K$ vs. M$_K$ relation. Filled squares indicate the 
data of de Grijs {\it et al.} \cite{deGrijs97b}, filled triangles the early-type
spirals of type $\le$ 1 of Peletier \& Balcells \cite{Peletier97}, and open circles
the early-type galaxies of Bower {\it et al.} \cite{Bower92a}. The thick lines 
connect some stellar population models, as discussed in the text. The thin
lines are least-squares fit to resp. the early and late-type galaxies.
{\bf b:} The same data, now as a function of morphological type.}
\end{figure}

We have plotted our CM relation in Fig.~5a, together with Bower's points. Also
drawn are two least-squares fits. The fit for the spirals was determined 
taking into account the errors in both directions. We find that the scatter
in $I-K$ amounts to 0.12 mag, but if one takes into account the observational
uncertainty, the scatter goes down to 0.07 mag, i.e. much smaller than was
published before. Fig.~1 shows that
\begin{itemize}
\item Spiral galaxies have a tight colour-magnitude relation, with a scatter
that is slightly larger than can be explained by observational uncertainties.
\item The relation has a different slope than the CM relation for ellipticals,
and there is an area between the two relation in which very few galaxies are
found.
\end{itemize}

We see two ways to explain these observational findings. The first is to revert
to the explanation of Tully {\it et al.} \cite{Tully82}, who state that there is
more star formation in spirals than in ellipticals. The blue light from
the stars affects $I-K$, although much less than e.g. $B-K$, and leaves 
M$_K$ almost unaffected. As an example we have drawn some two-burst stellar
population models using Bruzual \& Charlot 
\cite{Bruzual93} in Fig.~5a of 20 and 1 Gyr, 
where the younger component consists of a fraction x (50, 10 and 1\%) 
of the mass.
We see that for the smallest galaxies about 50\%  of the stars should be 
young, while for brighter galaxies this number should be 10\%, going down
more for the earliest type spirals. We could also take different stellar 
population models (for example 20 and 0.5 Gyr, exponentially decreasing or
continuous star formation), and find other numbers. The most interesting aspect
of this work is that we find that the 
scatter in the CM relation for spirals is so very small, and that there is
a gap between spirals and ellipticals. It means that the current star formation 
in a spiral galaxy is determined by its size, morphological type or luminosity, 
not by its environment, interactions or evolution. It also means that 
for example dwarf ellipticals are very different from dwarf irregular galaxies.

There is however another explanation, namely that in $I-K$ spirals have the same 
CM relation as ellipticals, but that the difference between the two relations 
is purely caused by stellar population gradients in the vertical direction 
in the spirals. In this case we still would have to explain why the scatter
between the spirals is so very small. Are vertical colour gradients important? 
At the moment we don't know very well
what the average  vertical colour gradients in spirals look like, but
we can make a quantitative estimate. For the sample of Peletier \& Balcells
\cite{Peletier97} the average $I-K$ colour gradient of the bulge was found to be
$\Delta(I-K)$/$\Delta(\log~r)$ = --0.19, with a RMS scatter of 0.09. 
The radial disc-gradients also are not zero: the average
$\Delta(I-K)$/(K-band scale length) here is --0.10, with a scatter of 0.09.
And furthermore bulges are slightly redder than discs (on the average  0.07
mag). Fisher {\it et al.} \cite{Fisher96} also show that  vertical 
metallicity gradients in bulges, as derived from the Mg$_2$ index, are much
larger than  radial disc gradients. In this paper the colour of the edge-on 
galaxies was determined between approx.
1 and 3 scale heights, and from the previous numbers it seems possible 
that the difference between the colour of the total galaxy and our colour
is as large as $\Delta(I-K)$ = 0.3 (or 0.40 in [Fe/H]), 
although a detailed study will have to
confirm whether this is really the case. The fact that the scatter between
spirals is so small indicates in this interpretation that all spirals
have similar colour gradients. Since vertical colour gradients would be 
larger than radial gradients, galaxies not seen edge-on, even without 
extinction, would create a larger scatter in the CM relation for spirals.

From the analysis presented in this paper we draw the following conclusions:
\begin{itemize}
\item We have determined a dust-free CM relation for spiral galaxies, by 
measuring $I-K$ colours in edge-on galaxies above the plane.
We find that the scatter in this relation is small and approximately as large as
can be accounted for by observational uncertainties. The slope of the IR 
CM-relation is larger for spirals than for elliptical galaxies.
\item We have two possible explanations. First, the difference could be 
caused by vertical colour gradients in spiral galaxies. In that case these
gradients should be similar from galaxy to galaxy, have an average size
of about 0.15 dex in [Fe/H] per scale height, and increase for later galaxy
types. Spirals and ellipticals would have the same colour-magnitude diagram,
indicating that the mass of a galaxy determines completely the old stellar 
populations, predicting a very low scatter in the IR Tully-Fisher law.
The other possibility would be that spirals and ellipticals have 
different CM relations. The difference would be caused by current star
formation, which has to be present in all spirals. The amount of 
current star formation would depend only on the galaxy luminosity, and 
not on environment.
\end{itemize}

\acknowledgements{We thank P.C. van der Kruit for useful discussions, and 
the conference organisers for organising a very pleasant and interesting meeting.}

\vfill
\end{document}